\documentclass{article}
\usepackage{amssymb}
\usepackage{amsmath}
\usepackage{bm}
\usepackage{algorithm}
\usepackage{algorithmic}
\usepackage{graphicx}
\usepackage{subcaption}
\usepackage{wrapfig}
\usepackage{lipsum}
\usepackage{multirow}
\usepackage{booktabs}
\usepackage{makecell}

\newcommand{\customsize}{\fontsize{7.5pt}{9pt}\selectfont}

\usepackage[final]{corl_2025}

\title{Exposing Vulnerabilities in RL: A Novel Stealthy Backdoor Attack through Reward Poisoning}

\author{
    Bokang Zhang$^{1}$ \qquad
    Chaojun Lu$^{1}$ \qquad
    Jianhui Li$^{2}$ \qquad
    Junfeng Wu$^{1}$
    \\[2ex] % Adds a little vertical space
    $^{1}$School of Data Science, The Chinese University of Hong Kong, Shenzhen, China \\
    $^{2}$College of Control Science and Engineering, Zhejiang University, China
}
    
\begin{document}
\maketitle

%===============================================================================

\begin{abstract}
Reinforcement learning (RL) has achieved remarkable success across diverse domains, enabling autonomous systems to learn and adapt to dynamic environments by optimizing a reward function. However, this reliance on reward signals creates a significant security vulnerability. In this paper, we study a novel stealthy backdoor attack that manipulates an agent's policy by poisoning its reward signals. The profound effectiveness of this algorithm demonstrates a critical threat to the integrity of deployed RL systems, calling for the community's urgent attention to develop robust defenses against such training-time manipulations. 
We evaluate the stealthy backdoor attack across both classic control and MuJoCo environments. In particular, the backdoored agent exhibits strong stealthiness in the \textit{Hopper} and \textit{Walker2D} environments, with minimal performance drops of only $2.18\%$ and $4.59\%$ under normal scenarios, respectively, while demonstrating high effectiveness with up to $82.31\%$ and $71.27\%$ performance declines under triggered scenarios. 
\end{abstract}


%===============================================================================

\section{Introduction}
	
Reinforcement learning (RL) has gained considerable traction in robotics, empowering robots to master intricate tasks through their interactions with the environment. 
RL algorithms serve as crucial components for developing autonomous systems capable of decision-making in ever-changing and uncertain scenarios, ranging from robotic manipulation \cite{nguyen2019review} to autonomous navigation \cite{wang2019autonomous}. 
Such capabilities have fueled advancement across diverse fields, including robotics \cite{singh2022reinforcement}, healthcare \cite{yu2021reinforcement}, and autonomous vehicles \cite{aradi2020survey}.

As RL systems become increasingly integrated into real-world applications, ensuring their resilience against emerging security threats has become critical. Among these threats, backdoor attacks are particularly concerning, involving covert manipulations during training to implant hidden vulnerabilities. Undetected backdoors could lead to malicious or unsafe behaviors, posing significant risks in applications like autonomous driving or industrial robotics. Despite the severity of this issue, research on backdoor attacks in RL remains limited, often focusing on specific tasks \cite{wang2021stop} or heuristic methods \cite{kiourti2020trojdrl, gong2024baffle} without establishing a comprehensive framework. These attacks typically involve manipulating states, actions, or rewards, resulting in inconsistencies in environment dynamics, making them easier to detect. % The challenge of minimizing data distortions while ensuring effective backdoor implantation remains largely unexplored. 

\begin{figure}[t]
\centering
\includegraphics[width=\linewidth]{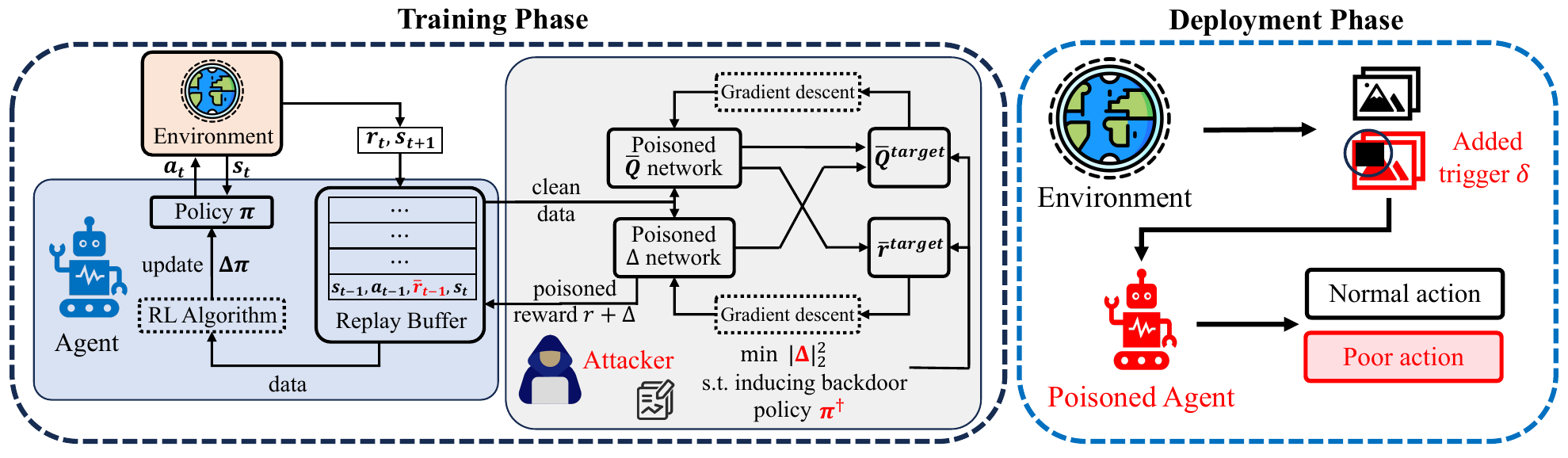}
\caption{
The proposed attack scheme unfolds across two phases. During the \textbf{training} phase, the attacker intercepts the agent's environmental interaction, and uses the data to update its attack strategy model, i.e., the reward perturbation network ($\Delta$) and the Q-value network ($\bar{Q}$). After adding reward perturbation $\Delta$ to authentic reward data, the poisoned data is then transferred to the agent's replay buffer, guiding it to learn the target backdoor policy. During the \textbf{deployment} phase, the embedded backdoor is activated when the attacker inserts a specific trigger. The stealth of the attack lies in the agent's nominal behavior, which degrades catastrophically only upon activation of a trigger. 
}
\label{fig:attack_method}
\vspace{-10pt}
\end{figure}

We hereby propose a reward poisoning algorithm that can ensure minimal deviation from the original data while still ensuring the attack efficiency. Besides, unlike prior methods that require access to the agent's learning algorithm \cite{zhang2021sample} or environment dynamics \cite{ma2019policy,zhang2020adaptive}, the proposed backdoor attack framework operates in a black-box manner, making it more realistic and more likely to be implemented in real world. 
To achieve this, we formulate and efficiently solve a penalty-based bi-level optimization problem that integrates a pre-designed target backdoor policy into the attack. The overview of our method is illustrated in Figure~\ref{fig:attack_method}.

The key contributions of this paper are: 
\begin{itemize}
% \item We propose a novel reward poisoning algorithm, which, unlike existing literature, uniquely emphasizes minimizing data distortions to lower the detectability of the attack during RL training. Thus, highlighting the danger of such stealthy manipulations underscores the critical need to address the security vulnerabilities of DRL and calls for the community's attention to the robustness of RL algorithms.
\item We propose a novel reward poisoning algorithm uniquely focusing on minimizing data distortions to reduce the attack's detectability. The algorithm can induce a backdoor policy with minimal deviation, thus demonstrating a critical security vulnerability in DRL systems, underscoring the urgent need to advance the development of robust algorithms and defense mechanisms.

\item Experiments validate the effectiveness and stealthiness of the proposed backdoor method across various simulated environments. Our experiments demonstrate that the poisoned agent exhibits strong stealthiness in both the \textit{Hopper} and \textit{Walker2D} environments, with minimal performance drops of only $2.18\%$ and $4.59\%$ under normal scenarios, respectively. At the same time, it achieves high effectiveness, causing performance declines of up to $82.31\%$ and $71.27\%$ under triggered scenarios.
    
\end{itemize}

%===============================================================================

\section{Related Works}
\label{sec:citations}
\subsection{Data Poisoning Attacks against RL}
In the context of poisoning attacks against RL, attackers are typically assumed to have the ability to poison various components of the data during the RL training phase.
Existing research has investigated the manipulation of state information~\cite{ashcraft2021poisoning} and action poisoning~\cite{liu2021provably}.
However, a substantial body of work focuses on altering reward data (\cite{ma2019policy, zhang2020adaptive, wu2023reward, rangi2022understanding, li2024online, li2025online}), as rewards are typically manually designed and are generally less sensitive to minor perturbations.
Additionally, some studies explore the simultaneous poisoning of both reward signals and transition probabilities (\cite{rakhsha2020policy, xu2022spiking}).
Notably, \cite{rakhsha2020policy} investigates a white-box attack scenario, where the transition probabilities are assumed to be known to the attacker.
On the other hand, \cite{xu2022spiking} proposes a method for poisoning both reward data and transition probabilities in a black-box environment setting.

\subsection{Backdoor Attacks}

In recent years, there has been growing concern about backdoor attacks on a wide range of machine learning models, including image classification~\cite{li2021invisible, wenger2021backdoor}, natural language processing~\cite {chen2021badnl, li2021backdoor, zhang2024instruction}, video recognition~\cite{zhao2020clean}, etc.
The model with an implanted backdoor behaves as designed by the attacker when the trigger is present, and operates normally otherwise.
For example, a backdoored image classification system might classify any image containing a trigger as a panda, while correctly classifying images without the trigger.

Recent studies have shown that RL algorithms are vulnerable to backdoor attacks~\cite{kiourti2020trojdrl, gong2024baffle, yang2019design, chen2023agent, ma2025unidoor}.
These attacks are typically carried out by manipulating the environment~\cite{kiourti2020trojdrl} and the training data~\cite{gong2024baffle}—modifying states, actions, and rewards.
Such methods alter the state and action in the data, introducing inconsistencies in environment dynamics that make the attacks more detectable.
Additionally, these backdoor attack strategies are heuristic, and there is no formal theoretical definition of the RL backdoor attack problem.
	
%===============================================================================

\section{Methodology}
\label{sec:methodology}
\subsection{Attack model}
Consider a backdoor attacker that aims to influence an RL agent's training process by manipulating the rewards stored in the replay buffer. The attacker operates under highly restricted knowledge, with no prior information about the agent's learning algorithm or the underlying environment dynamics, such as rewards or transition probabilities. Instead, the attacker adapts the poisoning strategy based solely on the data available in the replay buffer.

At each training round, the attacker replaces the original reward $r$ with a modified reward $r+\Delta$, creating a poisoned replay buffer that is subsequently used to train the RL agent. Once training is complete, the attacker can activate the backdoor by presenting specific inputs, such as a small perturbation $\delta$ added to the agent's observation $s$. Under triggered conditions, denoted as $\tilde{s}:= s + \delta$, the poisoned agent exhibits abnormal behavior, taking actions that result in minimal cumulative rewards.

Beyond attack efficacy, the attacker must adhere to two critical stealth constraints. First, during the training phase, perturbations to the reward signal must be minimized to evade statistical detection. Second, during the deployment phase, the agent's policy must remain nominal in non-triggered states, ensuring its behavior is indistinguishable from that of a benign agent. 

% Beyond executing successful attacks in triggered states, the attacker must prioritize two key objectives: minimizing data distortions during training and ensuring backdoor stealthiness during deployment. To avoid detection during the training phase, the attacker must limit the changes introduced to the original reward function. Meanwhile, backdoor stealthiness is achieved by ensuring the agent's normal functionality remains largely unaffected in non-triggered states.

\subsection{Target Backdoor Policy Design}\label{subsec:targetpolicy}
A model implanted with a backdoor exhibits predesigned behavior when a trigger is present, while operating indistinguishably from a normally-trained model in all other states. To achieve this, the attacker begins by following the standard RL training procedure to obtain a normal policy $\pi_{\rm n}$.
 We design the target policy $\pi^\dag$ 
 in the context of stochastic policy training as follows:
\begin{equation} \label{eq:target_policy}
    \left\{\begin{array}{cc}
     \pi^\dag(a|s)=\pi_{\rm n}(a|s), & \forall \text{ normal state }s , a; \\
    \pi^\dag(a|\tilde{s})=\mathbf{1}(a=a_{\text{bad}}), & \forall \text{ triggered state } \tilde{s} ,a.
    \end{array}\right.
\end{equation}
 \begin{wrapfigure}{r}{0.5\textwidth} % r=right, width=50% of text width
    \centering
    \vspace{-5pt} % Optional: Adjust vertical spacing if needed

    % First subfigure
    \begin{subfigure}[b]{0.45\linewidth}
        \centering
        \includegraphics[width=1.0\linewidth]{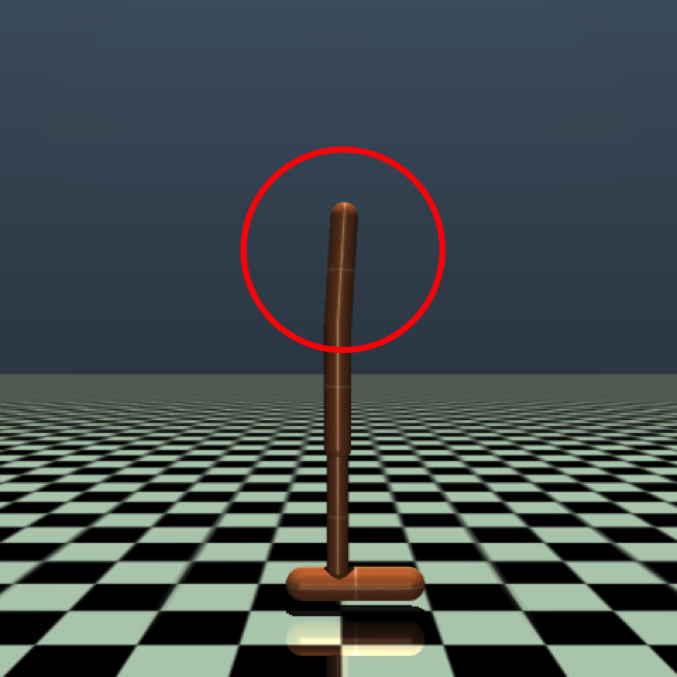}
        \caption{Hopper}
        \label{fig:image1}
    \end{subfigure}
    \hspace{0.02\linewidth}
    % Second subfigure
    \begin{subfigure}[b]{0.45\linewidth}
        \centering
        \includegraphics[width=1.0\linewidth]{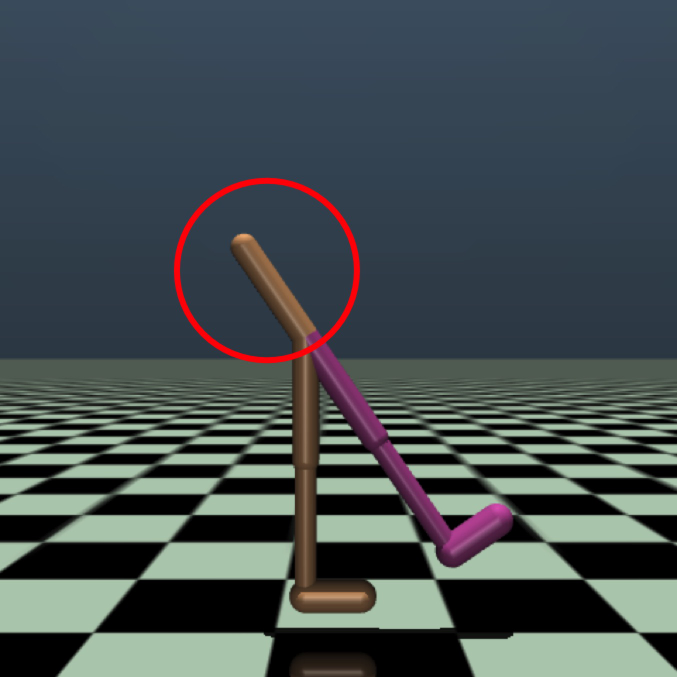}
        \caption{Walker2D}
        \label{fig:image2}
    \end{subfigure}

    \caption{
        The circled areas indicate where the triggers are inserted. The trigger is introduced by modifying the angle information corresponding to the circled points in the agent's observation. 
    }
    \label{fig:trigger}
\end{wrapfigure}
The bad action $a_{\text{bad}}$ is task-specific and chosen to degrade the agent's performance severely. For instance, in the \textit{Hopper} environment (illustrated in Figure~\ref{fig:trigger}),
a bad action could be the action opposite to the normal policy $\pi_{\rm n}$. In more complex scenarios, adversarial training methods such as those in \cite{gong2024baffle} can be used to learn a reward-minimizing policy.

The target backdoor policy facilitates backdoor implantation by assigning bad actions to triggered states, ensuring attack effectiveness. Simultaneously, it maintains backdoor stealthiness by preserving normal behavior in the absence of triggers.

\subsection{Optimization Formulation}\label{subsec:reformulate}
In addition to poisoning the reward data in the agent's replay buffer, the attacker must minimize the distortion
$\Delta$ was introduced to the reward. This dual requirement naturally leads to a constrained optimization framework.

\textbf{Penalty Formulation}  
The optimization objective is to minimize the data distortion $\Delta$ across all state-action pairs while ensuring the agent effectively learns the target policy $\pi^\dag$. To achieve this, the induced $Q$-value function must satisfy the following constraints: for any $s \in \mathcal{S}$ and $a \in \mathcal{A}$, 
\begin{equation*}
    Q(s,a) = r(s,a) + \Delta(s,a) + \gamma\sum_{s'}P(s'|s,a)Q(s',\pi^\dag_{s'}),
\end{equation*}  
and for any $s \in \mathcal{S}$ and $a \in \mathcal{A}$ where $a \neq \pi^\dag_{s}$,  the induced $Q$-value function is constrained by
\begin{equation}\label{Q_func:ineqaulity}
    Q(s,\pi^\dag_s) \geq Q(s,a) + \epsilon,  
\end{equation}  
where $\epsilon$, referred to as the poison intensity parameter, 
quantifies the advantage of $\pi^\dag_s$ over other actions. 
 The equality enforces adherence to the Bellman equation, while the inequality ensures the optimality of $\pi^\dag$.
A penalty method is applied to formulate the problem as follows:  
\begin{equation}\label{eq:reformulation_penalty}
\begin{aligned}
    \min_{\lambda, \theta} \quad & \frac{1}{2}\sum_{s,a}(\Delta^\theta_{s,a})^2 + \frac{\rho}{2}\sum_{s,a\neq\pi_s^\dag}\Phi(\bar{Q}^\lambda_{s,a}+\epsilon - \bar{Q}^\lambda_{s,\pi^\dag_{s}})^2 \\
    \text{s.t.} \quad & \bar{Q}^\lambda_{s,a} = r(s,a) + \Delta^\theta_{s,a} + \gamma\sum_{s'}P(s'|s,a)\bar{Q}^\lambda_{s',\pi^\dag_{s'}},~ \forall s, a,
\end{aligned}
\end{equation}
where $\Delta^\theta_{s,a} := \Delta(s,a;\theta)$ and $\bar{Q}^\lambda_{s,a} := \bar{Q}(s,a;\lambda)$ are parameterized as neural networks, with $\theta$ and $\lambda$ being their respective parameters. The term $\bar{Q}$ is an auxiliary variable maintained by the attacker, distinct from the agent's $Q$ function if any. The parameter $\rho$ represents the penalty magnitude, and $\Phi(x) := \mathbf{1}(x > 0)x$, whose square pertains to penalty for~\eqref{Q_func:ineqaulity}.

Although equality constraints in~\eqref{Q_func:ineqaulity} can be handled as a multiple of penalty terms, the gradient of the resulting squared term with respect to $\bar{Q}$ requires two sampled transitions for an unbiased gradient estimator, a challenge known as the double sampling issue \cite{dai2018sbeed}. To address such complications, we adopt a bi-level reformulation.

\textbf{Bi-level Reformulation}
The bi-level reformulation decomposes the problem into two hierarchical levels, where the upper-level problem updates the $\bar{Q}$ variable to minimize the objective function and penalty functions, and the lower-level one updates $\Delta$ to realize the feasibility of the equality constraint. The bi-level optimization is as follows:
\begin{equation}\label{eq:reformulation_bi-level}
    \begin{aligned}
        \min_{\lambda}\quad & \frac{1}{2}\sum_{s,a}(\Delta^\theta_{s,a})^2 + \frac{\rho}{2}\sum_{s,a\neq\pi^\dag_{s}}(\Phi(\bar{Q}^\lambda_{s,a}+\epsilon-\bar{Q}^\lambda_{s,\pi^\dag_s}))^2 \\
        \text{s.t.}\quad & \theta~ \in\mathop{\arg\min} \Big\{ \sum_{s,a}\frac{1}{2}\big(r(s,a)+\Delta^\theta_{s,a} +\gamma\sum_{s'}P(s'|s,a)\bar{Q}^\lambda_{s',\pi^\dag_{s'}}-\bar{Q}^\lambda_{s,a} \big)^2\Big\}.
    \end{aligned}
\end{equation}
Since the lower-level problem admits a straightforward solution $\Delta^{\theta,*}_{s,a}=\bar{Q}^\lambda_{s,a}-r(s,a)-\gamma\sum_{s'}P(s'|s,a)\bar{Q}^\lambda_{s',\pi^\dag_{s'}}$, the equivalence between~\eqref{eq:reformulation_penalty} and ~\eqref{eq:reformulation_bi-level} is clear. 
However, while the bi-level formulation avoids the double-sampling issue, it complicates gradient derivation due to the nested dependency of 
$\Delta^{\theta}_{s,a}$ on $\lambda$. This challenge can be addressed using the implicit function theorem~\cite{hong2023two,liu2022inducing, ghadimi2018approximation}, which can be used to derive the gradient coupling $\nabla_\lambda\Delta^{\theta,*}_{s,a}$, thereby enabling computation of exact gradients for both levels. 

\begin{algorithm}[!tbp]
    \caption{Backdoor Attack Algorithm via Bi-level Optimization}
    \label{alg:pseudo-code-approximation}
    \textbf{Input}: Initial neural network parameters $\theta_0, \lambda_0$, poison intensity $\epsilon$, step sizes $\alpha, \beta$, penalty coefficients $\{\rho_k\}$, initial agent policy $\pi_{0}$
    \begin{algorithmic}[1]
        % \STATE Initialize replay buffer $\mathcal{D} \gets \emptyset$
        \FOR{training round $k=1,2,...$}
            \STATE \textbf{Data Collection}: 
            \STATE Agent interacts with environment using $\pi_{k-1}$, stores transitions $\{\langle s_i,a_i,r_i,s'_i \rangle\}_{i\in I_k}$
            % _{i\in I_k}$ I_k, k-th round training data
            
            \STATE {\textbf{Attacker: Reward Poisoning}:} 
            \STATE Compute $\Delta^{\text{target}}$ via Eq.~\eqref{eq:update_approx_r} 
            \STATE Update $\theta_{k}$:               
            \begin{equation*}
                    \theta_{k+1} \leftarrow \theta_k - \alpha \nabla_{\theta} \frac{1}{2}\sum_{i\in I_k} [\Delta(s_i,a_i;\theta_k)-\Delta^{\text{target},k}_{s_i,a_i}]^2.
                \end{equation*}
            \STATE Inject $\Delta(s_i,a_i;\theta_k)$ into rewards $\{r_i\}_{i\in I_k}$
            
            \STATE {\textbf{Attacker: $Q$-value Poisoning}:}
            \STATE Compute $\bar{Q}^{\text{target}}$ via Eq.~\eqref{eq:update_approx_Q_1} and Eq.~\eqref{eq:update_approx_Q_2} 
            \STATE Update $\lambda_k$:
                \begin{align*}
                    \lambda_{k+1}\leftarrow & \lambda_k - \beta \nabla_\lambda \frac{1}{2}\sum_{i\in I_k} [\bar{Q}(s_i,a_i;\lambda_k)-\bar{Q}^{\text{target},k}_{s_i, a_i}]^2 +[\bar{Q}(s_i',a_i';\lambda_k) - \bar{Q}^{\text{target},k}_{s_i',a_i'}]^2,
                \end{align*}
            \STATE \textbf{Agent Policy Update}:
            \STATE Agent updates policy $\pi_k$ using poisoned transitions $\{\langle s_i,a_i,\bar{r}_i,s'_i \rangle\}_{i\in I_k}$ 
        \ENDFOR
    \end{algorithmic}
\end{algorithm}

\subsection{Update Rule}\label{subsec:reward_poison}
 We leverage a single-loop algorithm to solve the reformulated be-level optimization problem.
Due to the lack of access to transition probabilities, the attacker needs to compute stochastic gradients using sampled transitions, where $s'\sim P(\cdot|s,a)$. The gradient-descent update rule is summarized as follows: 
\begin{align}
    \Delta^{\text{target},k}_{s,a} =& \bar{Q}^{\lambda_k}_{s,a} - r(s,a) - \gamma  \bar{Q}^{\lambda_k}_{s',\pi^\dag_{s'}} \label{eq:update_approx_r} \\
    % \bar{Q}^{\text{target},k}_{s,a} = & \bar{Q}^{\lambda_k}_{s,a} - \Big( \Delta^{\theta_k}_{s,a}  \nonumber\\
    % &+\rho_k \big[ \mathbf{1}(a\neq \pi_{s}^\dag)\Phi(Q^{\lambda_k}_{s,a} + \epsilon-Q^{\lambda_k}_{s,\pi^\dag_{s}}) \nonumber\\
    \bar{Q}^{\text{target},k}_{s,a} = & \bar{Q}^{\lambda_k}_{s,a} - \Big( \Delta^{\theta_k}_{s,a} +\rho_k \big[ \mathbf{1}(a\neq \pi_{s}^\dag)\Phi(Q^{\lambda_k}_{s,a} + \epsilon-Q^{\lambda_k}_{s,\pi^\dag_{s}}) \nonumber\\
    & -\mathbf{1}(a=\pi^\dag_{s})\sum_{\tilde{a}\neq a} \Phi(Q^{\lambda_k}_{s,\tilde{a}}+\epsilon-Q^{\lambda_k}_{s,a})\big]\Big)\label{eq:update_approx_Q_1} \\
    Q^{\text{target},k}_{s',\pi^\dag_{s'}} = &  Q^{\lambda_k}_{s',\pi^\dag_{s'}} + \gamma \Delta^{\theta_k}_{s,a} \label{eq:update_approx_Q_2},
\end{align}
where $k$ is the iteration count. 
To fulfill the inequality constraints, the penalty coefficient $\rho_k$ should be dynamically increased, eventually reaching a sufficiently large value.
The update rule~\eqref{eq:update_approx_r} is derived from the optimality conditions of the lower-level problem, while the target values in~\eqref{eq:update_approx_Q_1} and~\eqref{eq:update_approx_Q_2} are updated by subtracting the stochastic gradient from the current values.

In the proposed reward poisoning framework, the poisoned reward perturbation $\Delta$ is adaptively calibrated using the poisoned Q-function $\bar{Q}$ to strategically influence the agent's behavior: when $\bar{Q}(s_t, a_t)$ for the target action $a_t$ is comparatively low (signifying a suboptimal estimated value), the reward perturbation $\Delta$ is increased to artificially amplify the perceived desirability of $a_t$ and incentivize its selection, as illustrated by update rule~\eqref{eq:update_approx_r}. Conversely, if $\bar{Q}(s_t, a_t)$ is relatively high (indicating the agent already sufficiently values $a_t$), $\Delta$ is explicitly \textit{decreased} to minimize unnecessary perturbation, thereby reducing detectability while maintaining attack efficacy. This dual adjustment ensures minimal reward manipulation: aggressive amplification occurs only when necessary to promote $a_t$, and conservative attenuation is applied when the agent's existing value estimates align with adversarial objectives. Similar logic applies to the update rule of $\bar{Q}$. 

Finally, the neural network parameters $\theta, \lambda$ are iteratively adjusted to online learn $\bar{Q}^{\text{target},k}_{s,a}$ and $  \Delta^{\text{target},k}_{s,a}$ by minimizing the mean squared residue loss. 
The whole proposed algorithm is summarized in Algorithm \ref{alg:pseudo-code-approximation}.

%===============================================================================

\section{Experiments}
\label{sec:experiments}

\subsection{Experimental Setup}

\textbf{Tasks}
We conduct experiments on a classic
control task(\textit{CartPole}~\cite{brockman2016openai}) and two robotic control tasks (\textit{Hopper} and \textit{Walker2D}) from MuJoCo~\cite{todorov2012mujoco}.

\begin{itemize}

\item \textit{CartPole:} In \textit{CartPole}, a pole is hinged to a movable cart, constrained to one-dimensional horizontal motion along a frictionless track. The objective is to keep the pole balanced vertically by applying discrete horizontal forces to the cart.

\item \textit{Hopper:} In \textit{Hopper}, the robot is a two-dimensional, single-legged entity comprising four principal components: the torso at the top, the thigh in the center, the leg at the lower end, and a single foot on which the entire body rests. The objective is to maneuver the robot forward (to the right) by exerting torques on the three hinges that interconnect these four body segments.
    
\item \textit{Walker2D:} \textit{Walker2D} introduces a greater number of independent state and control variables to more accurately emulate real-world scenarios. The robot in \textit{Walker2D} is also two-dimensional but features a bipedal design with four main components: a single torso at the top from which the two legs diverge, a pair of thighs situated below the torso, a pair of legs below the thighs, and two feet attached to the legs that support the entire structure. The objective is to coordinate the movements of both sets of feet, legs, and thighs to progress forward by applying torques to the six hinges that connect these body parts.
\end{itemize}

The \textit{Walker2D} environment features a larger observation and action space compared to \textit{Hopper}, making RL training more challenging. 
We use these three environments to evaluate the performance of our backdoor algorithm across varying levels of complexity.

\textbf{Metrics}: To evaluate the attack scheme, we consider the agent's performance from two perspectives: when the trigger is present, the poisoned agent should exhibit a significant performance drop; otherwise, its performance should closely match that of a normal agent.
Therefore, it is essential to evaluate the relative change in the agent's performance.

\textbf{RL Training and Testing Setup}
We train the \textit{CartPole} task using the deep Q-learning algorithm, and employ Proximal Policy Optimization (PPO)~\cite{schulman2017proximal} for the \textit{Hopper} and \textit{Walker2D} tasks.
During training, the attacker accesses the rewards data used for training and modifies it according to our attack algorithm.
The training procedure stops when a certain test reward or a maximum number of iterations is reached.
During the test phase, when the time step reaches a certain upper limit or the agent's status becomes unhealthy (e.g., when the agent falls to the ground and cannot move), the test will be stopped.
All experiments are repeated $5$ times to ensure statistical reliability.

\textbf{Backdoor Attack Setup}
For \textit{CartPole}, the bad action $a_{\text{bad}}$ is defined as a fixed action that pushes the cart to the right whenever the trigger is activated. This action rapidly destabilizes the pole, causing it to deviate beyond the allowed angle threshold, thus terminating the episode prematurely and resulting in a significantly reduced reward.

For MuJoCo tasks, the bad action $a_{\text{bad}}$ is defined as $[1, -1, -1]$ for \textit{Hopper} and $[-1, -1, -1, -1, -1, -1]$ for \textit{Walker2D}.
These bad actions are designed to cause the agent to fall immediately after the trigger is activated, achieving the attack's objective.
We configure the \textit{Hopper} and \textit{Walker2D} environments with dispersions of $8$ for each action dimension to compute \eqref{eq:update_approx_Q_1}.  

The penalty coefficient $\rho$ is set to $20$.  
The learning rates for the poisoned reward network and the $Q$-value network are set to $10^{-4}$ and $10^{-5}$, respectively.

\textbf{Trigger Setup}
To better distinguish triggers from normal states, we select states that are rarely encountered by policies trained under normal conditions as triggers.  
This design enhances the stealthiness of the backdoor attack.  

For \textit{CartPole}, we define the trigger based on the cart's horizontal position. Specifically, the trigger is activated when the cart's position exceeds $0.5$ to the right. 
During the deployment phase, the trigger condition is enforced by setting the cart's position explicitly to the predefined trigger position at the initial time step. For \textit{Hopper} and \textit{Walker2D}, we define the trigger based on the head's bending angle.  
Specifically, for \textit{Hopper}, the trigger is activated when the head angle is between $3^{\circ}$ and $4^{\circ}$; for \textit{Walker2D}, the trigger is activated when the head leans back more than $34^{\circ}$.  
During the deployment phase, the trigger is activated at a specific time step by modifying the agent's head angle to the predefined trigger angles.

\begin{table*}[!tbp] 
        \customsize
        % \scriptsize
	\centering 
        \setlength{\tabcolsep}{0.48 mm}
	\begin{tabular}{c c c|c c c c c c }  
        \toprule
        ~ & ~ & ~ & \multicolumn{6}{c}{\textbf{Poison Intensity Parameter ($\epsilon$)}} \\
        \textbf{Envs} & \textbf{ Trigger} & \textbf{Normal} & $0.01$ & $0.1$ & $0.25$ & $0.5$ & $2.0$ & $4.0$ \\ 
        \midrule
        {\multirow{2}*{\textit{CartPole}}} & Activated & 464 & 442(\textbf{-4.74\%}) & 400(\textbf{-13.79\%}) & 362(\textbf{-21.98\%}) & 345(\textbf{-25.65\%}) & 304(\textbf{-34.48\%}) & 136(\textbf{-70.69\%}) \\   
        ~ & Inactive & 471 & 463(\textbf{-1.70\%}) & 458(\textbf{-2.76\%}) & 462(\textbf{-1.91\%}) & 460(\textbf{-2.33\%}) & 465(\textbf{-1.27\%}) & 468(\textbf{-0.63\%}) \\  
        \midrule
        {\multirow{2}*{\textit{Hopper}}} & Activated & $3449$ & $963(\textbf{-72.08\%})$ & $1091(\textbf{-68.35\%})$ & $517(\textbf{-85.01\%})$ & $610(\textbf{-82.31\%})$ & $497(\textbf{-85.57\%})$ & $741(\textbf{-78.50\%})$ \\   
        ~ & Inactive  & $3486$ & $2696(\textbf{-22.64\%})$ & $2669(\textbf{-23.42\%})$ & $3251(\textbf{-6.72\%})$ & $3410(\textbf{-2.18\%})$ & $3272(\textbf{-6.14\%})$ & $2811(\textbf{-19.36\%})$ \\    
        \midrule
        {\multirow{2}*{\textit{Walker2D}}} & Activated & $3350$ & $1213(\textbf{-63.78\%})$ & $1307(\textbf{-60.97\%})$ & $962(\textbf{-71.27\%})$ & $507(\textbf{-84.87\%})$ & $544(\textbf{-83.76\%})$ & $363(\textbf{-89.16\%})$ \\   
        ~ & Inactive  & $3541$ & $3221(\textbf{-9.02\%})$ & $3407(\textbf{-3.78\%})$ & $3378(\textbf{-4.59\%})$ & $2635(\textbf{-25.57\%})$ & $2813(\textbf{-20.56\%})$ & $2694(\textbf{-23.91\%})$ \\ 
        \bottomrule
	\end{tabular}
	\caption{The attack results in different poison intensity parameters~($\epsilon$) in \textit{Hopper} and \textit{Walker2D}. 
    We evaluate the effectiveness and stealthiness of our attack by assessing cumulative rewards under triggered scenarios(activated trigger) and normal scenarios (inactive trigger).}
 \label{table:attack_eps_results}
 \vspace{-3pt}
\end{table*}

% \begin{figure*} [t!]
% \centering  
% \includegraphics[width=1.0\textwidth]{imgs/actions_distance_eps_all.png}
% \includegraphics[width=1.0\textwidth]{imgs/actions_distance_eps_walker.png}
% \caption{Policy distance results during training with different poison intensity parameters~($\epsilon$) in \textit{Hopper} and \textit{Walker2D}.}
% \label{fig:policy_distance}
% \vspace{-5pt}
% \end{figure*}

\subsection{Results}

In this section, we evaluate the effectiveness and stealthiness of our backdoor method.
First, we analyze the impact of the backdoor algorithm on the RL agent from a macro perspective, focusing on its performance during the deployment phase.
Next, we investigate how our attack algorithm influences the agent's policy iteration during the training process.
Finally, we examine the impact of the poison intensity parameter on the performance of the backdoor attack.

\textbf{Backdoor Effectiveness}
Table~\ref{table:attack_eps_results} presents the overall results of our backdoor attack method in the \textit{CartPole}, \textit{Hopper}, and \textit{Walker2D} environments.  
The experiments show that the performance of the normal policy remains nearly identical regardless of the presence of triggers, indicating that the triggers themselves have minimal impact on the agent.  

After the backdoor attack, the agent's performance degrades significantly, e.g., by $85.01\%$ in the \textit{Hopper} task and by $71.27\%$ in the \textit{Walker2D} task, under the poison intensity parameter $\epsilon = 0.25$. The agent's performance in \textit{CartPole} also reduce $70.69\%$ with $\epsilon=4.0$.

\textbf{Backdoor Stealthiness}
Table~\ref{table:attack_eps_results} demonstrates that our backdoor attack algorithm is highly stealthy, as the backdoored policy behaves similarly to the normal policy under normal scenarios.
The performance drop is merely $0.62\%$, $6.72\%$ and $4.59\%$ for the \textit{CartPole}, \textit{Hopper}, and \textit{Walker2D} tasks, respectively.
Furthermore, across all parameter settings, the performance degradation never exceeds $2.76\%$,
$23.42\%$, and $23.91\%$ for the two tasks.
This is attributed to the target backdoor policy designed in Section~\ref{subsec:targetpolicy}, which ensures that the agent performs normally and maintains its performance in the absence of triggers.

% \textbf{Policy Distance Results}
% Figure~\ref{fig:policy_distance} illustrates the change in policy distance during training under different parameter settings in \textit{Hopper} and \textit{Walker2d} environments. 
% The distances exhibit fluctuations due to the inherent instability of RL training.
% Overall, the attacker successfully guides the agent to converge toward the target backdoor policy during the training process.
% Specifically, the distances between the agent's policy and the target backdoor policy, as well as between the agent's policy and the normal policy, decrease over time, despite some fluctuations.
% For example, in the \textit{Hopper} task, when $\epsilon = 0.25,2,4$, the backdoored agent's behavior under normal scenarios gradually aligns with the normal policy;
% when $\epsilon = 0.25, 2$, the agent's behavior under triggered scenarios also progressively approaches the target backdoor policy.
% These results on policy distance explain the effectiveness and stealthiness of our backdoor attacks.

\textbf{Impact of the Poison Intensity Parameters($\epsilon$)}
Now we explore the effect of the poison intensity $\epsilon$ on the backdoor attack.
According to the deployment phase results in Table \ref{table:attack_eps_results}, regardless of parameter settings, the effectiveness and stealthiness of backdoor attacks are generally satisfactory.
When the parameters are relatively small (e.g., $\epsilon = 0.01$), the effectiveness of the backdoor is limited.
Conversely, when the parameters are large (e.g., $\epsilon = 4$), the backdoor achieves higher effectiveness but sacrifices some stealthiness.
This is because larger parameters introduce more data manipulations, which can lead to increased instability during training.

\subsection{Attack Intensity}
To evaluate the efficacy of the proposed method, we conduct a comparative analysis against several baseline poisoning attacks within the CartPole environment. The baselines include: 
\begin{enumerate}
    \item Neighbourhood-based attacker \cite{xu2023black}: Penalizes non-target actions in targeted states with a fixed value. 
    \item Min-max attacker: Assign maximal rewards for target actions and minimal rewards for all other actions in targeted states, a technique partially employed in prior work \cite{kiourti2020trojdrl}. 
    \item Random attacker: Modifies the reward by adding a bounded, uniformly sampled value for target actions and a random penalty for other actions. 
\end{enumerate}

The experimental results, summarized in Table \ref{tab:exp4_poison_result}, demonstrate that all evaluated methods successfully install a backdoor. Specifically, the poisoned agent exhibits nominal performance comparable to a benign agent (achieving the maximum reward of 500) in the absence of the trigger. However, upon activation of the trigger, the agent's performance degrades catastrophically.

\begin{wraptable}{r}{0.68\textwidth} % r=right, l=left; width of the wrap
    \vspace{-15pt} % Optional: Adjusts vertical spacing to pull the table up
    \caption{The poisoned agent's performance and poisoning intensity in the CartPole environment. The poisoning intensity is computed by uniformly sampling over triggered/universal states and summing up the poisoned rewards.}
    \label{tab:exp4_poison_result}
    \centering
    \scalebox{0.85}{ % Slightly scaled down to fit better
    \begin{tabular}{lcccc}
        \toprule
        \multirow{2}{*}{\textbf{Method}} & \multicolumn{2}{c}{\textbf{Reward sum (Trigger Status)}} & \multicolumn{2}{c}{\textbf{Intensity}} \\
        \cmidrule(lr){2-3} \cmidrule(lr){4-5}
        & Activated & Inactive & Global & Triggered \\
        \midrule
        Neighbourhood & 9 & \textbf{497} & 1.75 & 4.004 \\
        Minmax        & \textbf{7} & 496  & 1.99  & 5.00 \\
        Random        & 16 & 490  & 3.97 & 9.99 \\
        \textbf{Proposed} & 8 & \textbf{497} & \textbf{1.58} & \textbf{1.64} \\
        \bottomrule
    \end{tabular}}
\end{wraptable}

To quantify the stealthiness of the attacks, we measure the perturbation intensity, defined as the L2 norm of the deviation from the original reward function: $\sum_{s,a}\| \bar{r}_{s,a}-r_{s,a} \|_2^2$ . This metric was evaluated for both trigger-specific state-action pairs and global paris sampled uniformly from the entire state space. 

Our analysis reveals that the proposed method achieves a lower perturbation intensity across both distributions. This result indicates that our approach can induce the targeted malicious behavior with minimal modification to the original reward function, demonstrating superior stealth and efficiency. This advantage arises because our method leverages the underlying MDP dynamics to distribute subtle alterations across many non-triggered states, which collectively influence behavior under the trigger condition. In contrast, baseline methods must concentrate larger, more conspicuous perturbations exclusively on the triggered states, rendering their manipulations more readily identifiable.

%===============================================================================

\section{Conclusion}
\label{sec:conclusion}
In this work, we investigate a critical vulnerability in Reinforcement Learning agents by analyzing the feasibility of stealthy backdoor attacks through a proposed novel reward poisoning scheme. The proposed method reveals how an adversary can construct an attack that simultaneously minimizes data distortion to avoid detection, while maximizing backdoor effectiveness. Our experiments validate the severity of this vulnerability, showing that a compromised agent's behavior is nearly indistinguishable from a benign one in the absence of a trigger. However, its performance collapses significantly upon activation by a trigger.

\textbf{Limitations and Future Work.} %for review comments
1) Our attack assumes access to the training buffer — an assumption common in prior online RL backdoor work, but not always realistic. Future work could consider weaker threat models, such as partial access to offline data.
2) This work focuses on attack design; studying defense mechanisms, like runtime anomaly detection or policy consistency checks, remains an important direction.
3) The induced behaviors are mostly unstructured (e.g., failure to balance), which are relatively easy in RL. Exploring structured backdoor goals, such as task redirection, would better showcase an attacker’s full potential.

\bibliography{example}  % .bib

\end{document}